\documentstyle[11pt,paspconf,psfig]{article}

\def\lya{\mbox {Ly$\alpha$~}}
\def\IZw18{I~Zw~18}
\def\m82{M82}

\def\msun{\mbox {${\rm ~M_\odot}$}}
\def\zsun{\mbox {${\rm ~Z_{\odot}}$}}

\def\angs{\mbox {~\AA}}
\def\Ha{\mbox {H$\alpha$~}}

\def\o3hb{[OIII]$\lambda5007$~/~H$\beta$~}
\def\O1ha{[OI]$\lambda6300$~/~H$\alpha$~}

\def\s2ha{[SII]$\lambda\lambda6717,31$~/~H$\alpha$~}
\def\2z2{HeII~$\lambda4686$~}
\def\z7{[NII]~$\lambda6583$ }
\def\N2{[NII]~$\lambda6583$~/~H$\alpha$~}
\def\16z2{[SII]~$\lambda\lambda6717, 6731$ }

\def\n{NGC~}
\def\asec{\ifmmode {'' }\else $''~$\fi}  
\def\amin{\ifmmode {' }\else $'~$\fi}    
\def\arcsper{\ifmmode \rlap.{'' }\else $\rlap{.}'' $\fi} 
\def\arcmper{\ifmmode \rlap.{' }\else $\rlap{.}' $\fi} 
\def\sles{\lower2pt\hbox{$\buildrel {\scriptstyle <}
   \over {\scriptstyle\sim}$}} 
\def\sgreat{\lower2pt\hbox{$\buildrel {\scriptstyle >}
    \over {\scriptstyle\sim}$}} 
\def\kms{~km~s$^{-1}$~}

\def\flux{~ergs~s$^{-1}$~cm$^{-2}$}
\def\cm3{~cm$^{-3}$}
\def\cm3{~cm$^{-2}$}

\def\et{{\rm et\thinspace al.}\ }   

\def\apj{ApJ}
\def\apjs{ApJS}

\def\aj{AJ}
\def\mn{MNRAS}

\def\aa{A\&A}

\begin{document}

\title{UV Diagnostics of the Interaction between
Starburst Galaxies and the Intergalactic Medium
}
\author{Crystal L. Martin\altaffilmark{1}}
\affil{Space Telescope Science Institute, Baltimore, MD 21218}
\altaffiltext{1}{Hubble Fellow}

\begin{abstract}
Both galaxies and the intergalactic medium evolve dramatically 
between $z \approx 2$ and the present. These changes are coupled via
galactic winds, cloud accretion, and ionizing radiation.  Measurements
of these interactions are critical for building a physical understanding
of galaxy assembly.  This paper reviews the role of starburst galaxies
in heating, enriching, and ionizing the intergalactic medium.  The strategies 
used to address these problems rely heavily on access to the rest-frame 
ultraviolet (UV) spectrum.  Rather modest gains in UV sensitivity are shown
to produce enormous gains in our ability to follow the very low-density gas 
dispersed by galactic winds and galaxy interactions.  This leverage results
from the sensitivity of absorption lines to low ionic columns and the very
steep rise in the areal density of quasars toward fainter magnitudes.
\end{abstract}


\section{Introduction -- The Assembly of Galaxies}
Deep imaging with the Hubble Space Telescope (HST) has been an important
catalyst for measuring the cosmic star formation history (Madau \et 1996).
Although corrections for extinction remain controversial, the basic results
support predictions for the hierarchical growth of galaxies remarkably well
(Baugh, Cole, Frenk, \& Lacey 1998 and references therein).  
The star formation history alone, however, is hardly a sufficient test of this
paradigm.  One needs to measure the assembly of mass, the dispersal of metals,
and the evolution of the angular momentum in galaxies.  

To build this physical understanding of galaxy evolution, spectra of 
thousands of galaxies will be obtained with 6-10~m ground-based
telescopes and the Next Generation Space Telescope (NGST).
Infrared spectra of galaxies at redshifts $z \ge 0.5$ will contain
strong rest-frame-optical emission lines.  Star formation rates, metal
abundances, and rotation curves will be derived using the same diagnostics
developed for nearby galaxies.  To avoid the airglow lines of the night sky,
many ground-based spectra of $z \ge 1$ galaxies will be obtained at optical 
wavelengths, and interpretation will rely on diagnostics in the less familiar
ultraviolet spectral domain.

The evolution of galaxies cannot be separated, however, 
from the evolution of the intergalactic medium (IGM).  In particular, gas 
inflow and outflow from galaxies affects the reservoir available for star 
formation.  Metals ejected into the IGM should be counted when computing
the integral of the star formation rate over cosmic time.  And starburst
galaxies may play a role in the ionization of the IGM. Even at the present 
epoch, galaxies are not thought to be the primary reservoir of baryons  
(e.g., Cen \& Ostriker 1998), so their interaction with this encompassing 
environment is clearly of fundamental interest for their origin, structure,
and evolution.

This paper reviews the role of ultraviolet spectroscopy in building
an understanding of galaxy formation and evolution. Applications directly
related to the starburst -- IGM interaction are emphasized, and 
the required sensitivity is explored.  Relatively little
will be said about important applications to the stellar population and
dust content of galaxies.

\section{UV Spectral Diagnostics}

Starburst galaxies have intrinsically blue spectral energy 
distributions. The Lyman break at 912 \angs\ is the strongest
continuum feature in the entire spectral energy distribution (SED).
Most of the resonance lines of common ions also lie in the ultraviolet 
spectral domain.  The absorption and emission lines in starburst spectra 
have three distinct origins:  stellar photospheres, stellar winds, and 
the interstellar medium/halo (Gonzalez-Delgado \et 1998) and
are sensitive probes of the ISM/halo over a wide range of ionization states.  
Measurements of ion columns and velocity dispersions are not limited 
to star-forming regions.

\subsection{Continuum Emission}

\vspace*{2.5in}
{Figure~1. Starburst oxygen abundance by number (the solar value is
8.93) vs. the spectral slope in the UV, where $F_{\lambda} \propto
\lambda^{\beta}$ (Heckman \et 1998).  }   \\  
 

The International Ultraviolet Explorer satellite obtained spectra of 
over 100 nearby, star-forming galaxies (Kinney \et 1993). Figure~1 shows 
that the spectral index of the starburst SED, $F_{\lambda} \propto 
\lambda^{\beta}$, is systematically redder in higher metallicity galaxies
(Heckman \et 1998).  In metal-rich, dusty starbursts, the stellar continuum 
is intrinsically blue, but much of the UV is absorbed
and reradiated at longer wavelengths.  The continua of metal-poor galaxies 
in the local universe do remain quite blue.  Figure~1 shows that the 
observed spectral index approaches the intrinsic value $\beta \sim -2$ in 
galaxies with $\log O/H \approx -4$.  Many of these dwarf galaxies are young
and only a small fraction of their total gas mass has been turned 
into stars. 

The oldest stars in the universe, some of which might have formed in very 
early starbursts, are also studied in the UV spectral domain. The 
continuum radiation from E galaxies and bulges of spiral galaxies 
plummets around 4000\angs, but shows a strong upturn shortward of 2700\angs\
(Bertola \et 1982, Figure 1)  due to horizontal branch stars. 
In nearby galaxies, this population may be resolved with HST/STIS. 
Post extreme horizontal branch stars, a small fraction of the population,
have been resolved in M32 and M31 with HST/FOC (Brown \et 1998).

\subsection{Starburst Sightlines}

Spectra with HST/GHRS of UV-bright star clusters in several galaxies 
were obtained (Kunth \et 1998; Heckman \& Leitherer 1997). The narrow, 
low-ionization lines arise from absorption in the ISM of these starburst 
galaxies and are sometimes blueshifted
with respect to the photospheric lines. Figure~2 shows 
Si~II and Si~III absorption in \n1705, a postburst dwarf galaxy,
at a velocity 100\kms less
than the photospheric C~III absorption. The optical emission-line radiation
splits into two components, redshifted and blueshifted by similar
amounts (Meurer \et 1992).
The UV measurements confirm that this shell of gas is expanding, rather than
infalling and show that a cooler gas phase is participating in 
the outflow.

\vspace*{2.5in}
{Figure~2. HST GHRS spectrum of a dwarf starburst galaxy
NGC~1705 from the study of Heckman \& Leitherer (1997).  } \\

The interstellar absorption lines in \n1705 are wider than the corresponding
Galactic absorption lines.  The intrinsic width, $\sim 120$\kms, probably
represents the macroscopic motion of the absorbing clouds.  In general, 
the broadening of interstellar lines, unlike stellar wind lines, 
is not directly correlated with the ionic column 
(Heckman \et 1998).  The width seems to be more closely related
to the strength of the burst and the total mechanical energy available.

Higher resolution GHRS spectra show \lya\ in emission in 4 of 8 H~II galaxies
with blueshifted interstellar absorption lines (Kunth \et 1998). Although \lya\
is expected to be produced in abundance, virtually none is expected to escape 
a static, homogeneous medium. Resonant scattering of the \lya\ photons greatly
increases the probability for absorption by dust in a static medium 
(Charlot \& Fall 1991). However, since the cross section drops abruptly 
longward of  the resonance, \lya\  backscattered from the receding side 
of an expanding H~I shell could escape (Ahn \& Lee 1998).  Indeed, Kunth \et 
suggest that the detected emission is redshifted and find \lya\  absorption 
below the systemic velocity -- i.e. a P~Cygni-like  line profile.  
The velocity structure of the ISM seems to be a very important factor
for determining the detectability of \lya\ emission.  

A typical high-redshift, Lyman break galaxy shows relatively weak  
stellar wind lines, blueshifted lines from ions in lower ionization states,
and a P-Cygni-like \lya\ profile (Lowenthal \et 1997).  
Outflows could therefore be prevalent in high-redshift galaxies. However,
deriving their properties from these line 
measurements clearly requires a local calibration.

\subsection{Halo Sightlines}

Galactic winds must be transient phenomena. Starbursts last of order
10~Myr, and the supernova rate plummets after 40~Myr. Following 
the ejected material well beyond the stage where it is illuminated by the
starburst is critical for determining its fate. Absorption-line studies using 
QSO sightlines are the best way to follow 
this material with low column density and unknown ionization state.
They also provide a means to measure the rotational speed of the neutral 
medium at column densities several orders of magnitude below the 
$\sim 10^{19}$~cm$^{-2}$ limit of 21-cm emission.  Such constraints on the 
extent of the dark matter halo are needed to constrain
models of the escape velocity.  Few sightlines are currently accessible, 
although two bright quasars are projected near the starburst galaxy NGC~520.
At a separation of $24h^{-1}$~kpc, Mg~II absorption is detected; and
a possible detection at $53h^{-1}$~kpc is reported (Norman \et 1996). 

Testing all galaxies for the presence of absorbing halos would provide an 
important link to the ISM in high redshift galaxies.  In QSO spectra, 
Mg~II $\lambda\lambda 2796,2803$ absorption lines at $0.2 < z < 1$
are associated with galaxies at separations $< 40h^{-1}$~kpc from the 
sightline (Steidel \et 1994).  Equivalent widths, $W_{\lambda} > 0.3$\angs, 
correspond to $N_{\rm HI} \ge 2 \times 10^{17}$~cm$^{-2}$.  
Sightlines through disks and halos
of nearby galaxies typically show no absorption at separations 
$\ge\ 31h^{-1}$~kpc to even higher sensitivities, $W_{\lambda 2796} = 0.04$ 
to 0.09\angs\ (Bowen \et 1995).  Figure 3 illustrates how similar 
these results are. 
The sharp cutoff could be an ionization effect since Mg~II disappears
as the H~I becomes optically thin at the Lyman limit (Steidel \&
Sargent 1992). 

\vspace*{2.5in}
{Figure~3. Separation of QSO sightline versus galactic magnitude from
the study of Bowen, Blades, \& Pettini (1995).  Filled symbols represent
detections of Mg~II absorption.  The dashed line represents the gas
cross section in moderate redshift galaxies.  } 

\subsection{Metal Enrichment}

The metallicity of star-forming galaxies has typically been derived
from the emission-line ratios of H~II regions.  Although their optical spectra
contain prominent lines of O, N, and S, access to the ultraviolet spectrum
is required to measure the C abundance.  The C/O ratio, typically  
C~III]$\lambda 1909$/[O~III]$\lambda 5007$, provides a
measure of the time elapsed since the last burst. The stars producing
most of the C live much longer than those synthesizing O.
The relatively high ratio of C/O in I~Zw18, for example, 
has been used to argue that some stars formed
in this  notoriously oxygen-poor galaxy
prior to the current burst (Garnett \et 1997).  For higher redshift galaxies,
more closely spaced lines would provide a more accessible
pollution index.  Initial efforts to calibrate the C~II] 2326 to [O~II] 
3727,29 ratio as a measure of the C/O ratio show promise, but
C measurements are needed for more local galaxies 
(Kobulnicky \& Skillman 1998). 

Ultraviolet absorption-line studies are potentially a sensitive
probe of abundance ratios.  The equivalent width of stellar-wind lines may 
be a good metallicity indicator (Heckman \et 1998).  Interstellar 
absorption provides an unique opportunity to measure abundance ratios in 
infalling gas, tidal ejecta, and galactic winds.  But sightlines toward 
the starburst region typically have high H~I column densities, and
many interstellar lines are optically thick (Heckman \et 1998)
and insensitive to abundance (Fig. 4). Little has been done 
conclusively at current sensitivities  (Pettini \& Lipman 1995).

\vspace*{4.0in}
{Figure~4. Theoretical fits to the O~I$\lambda 1302$ profile of
I~Zw~18 as shown by Pettini and Lipman (1995).  }   

\subsection{Do Ionizing Photons Escape from Galaxies?}

Starburst galaxies could be substantial contributors to
the metagalactic ionizing radiation field at $z \ge\ 3$
(Miralda-Escud\'e \& Ostriker 1990; Madau \& Shull 1996).
The observational evidence for escaping Lyman continuum radiation 
is not yet definitive, however.  An H~I column of only $1.6 \times 10^
{17}$~cm$^{-2}$ is sufficient to absorb a 1 Rydberg photon.  Yet, in 
nearby galaxies \Ha emission from recombining gas is seen 
at distances $\sim 1$~kpc from H~II regions; 
and the spectral gradient is often consistent with the  dilution of photons 
leaking out of H~II complexes (Martin 1997).  
This large propagation distance must be closely connected to the
geometry of the ISM (Dove \& Shull 1994). The excitation mechanism 
at many times this distance, where extremely low surface brightness 
\Ha emission has been detected, will not be determined unambiguously
until spectra can be obtained (Tufte \& Reynolds 1998; 
Bland-Hawthorn \et 1997).

\vspace*{2.5in}
{Figure~5. Total number of Lyman continuum photons vs luminosity at
900\angs\ for three starburst models from the study of 
Leitherer \et 1995.  }  \\
 
Measurement of the Lyman continuum flux shortward of the Lyman limit
is a more direct approach.  The recombination rate measured from the 
Balmer lines puts a lower limit on the production rate of ionizing photons.
The ratio of this production rate to the luminosity density at 900\angs\ is 
relatively insensitive to the starburst parameters as illustrated in Figure~5.
Upper limits on $L_{\lambda}^{900}$ therefore
provide upper limits on the escape fraction.
The limits derived from HUT spectra of 4 nearby starbursts
range from 3\% (Leitherer \et 1995) to as much as 57\% (Hurwitz \et 1997).
The difference arises from the treatment of 
the complex gas-phase absorption of the Milky Way.

\section{Science Drivers}

Measurements of ionic columns and the Lyman break are critical
for understanding the impact of starburst galaxies on the IGM and
the influence of environment on the starburst phenomenon.  These types
of studies require access to the ultraviolet spectral domain.  To answer
questions pertaining to the ubiquity of halos, hundreds of galaxies should
be probed.  Since the gas may not recombine as fast as it cools, higher
ionization states need to be probed as well as those in the H~I medium.
The most serious limitation to the interpretation is the ambiguity between 
absorption from relict winds and tidally stripped gas.  
Access to large samples of galaxies and multiple sightlines 
per galaxy are needed will resolve this issue.
In this section, I discuss the limits of absorption-line studies
with the HST instruments STIS and COS; sensitivity to the Lyman-break is
compared at several redshifts.  The goal is to determine how large a 
change in sensitivity would be required to enable a new genre of research
programs in this area.

\subsection{Availability of Sightlines}
Absorption lines are only powerful probes if bright quasars, star clusters,
gamma ray bursts, supernovae, etc. can be found sufficiently close to the
galaxy under study. Star clusters have several disadvantages.   Studies are
limited to the subset of starburst galaxies with low reddening,
and absorption from the ISM in the starburst region can be confused 
with that from an extended halo.  Quasar sightlines provide
 unambiguous probes of the extended halo.  Figure~6 illustrates the
growth in the number of quasars per square degree toward fainter
magnitude.  A rather modest gain in sensitivity, factor
of 2.8, provides access to an order of magnitude more sightlines!

\hbox{  \psfig{file=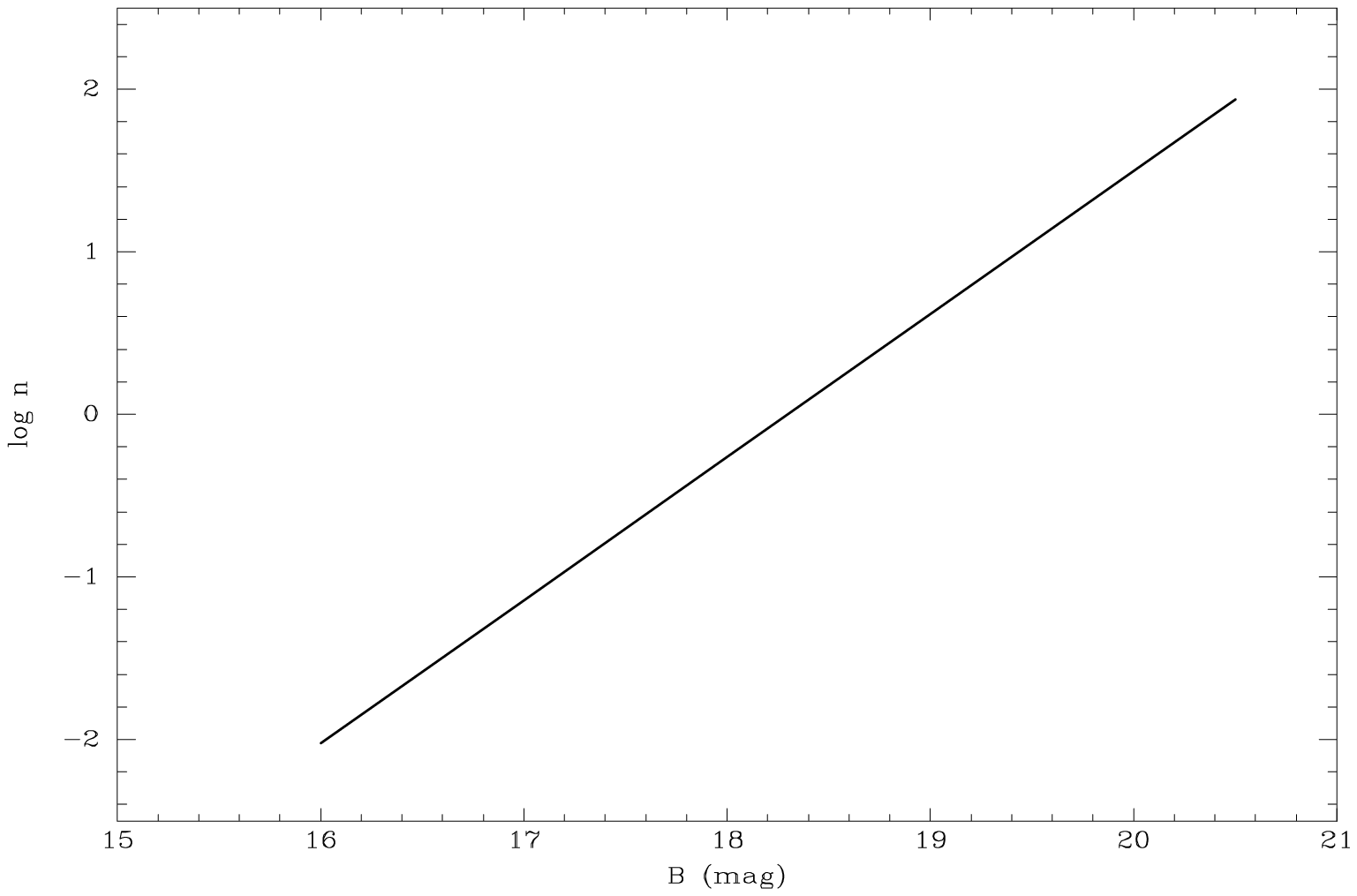,angle=0,height=2.5in} \hfill}
{Figure~6. Number of QSOs per unit area vs apparent magnitude based on
model of Weedman (1986).}
\vskip\the\baselineskip\

Most of the moderate-redshift metal line systems have been associated
with bright, $L \sim L^*$, galaxies.   When fainter galaxies are detected
the separation from the sightline tends to be lower (Steidel 1995;
Chen \et 1998).  Since material is clearly being ejected from some dwarf 
galaxies, the lack of dwarf galaxy absorption line systems is quite 
interesting (McLin, Giroux, \& Stocke 1998).
Suppose gas streams out of a ruptured superbubble at the sound speed,
$\sim 300$\kms.  Between bursts, perhaps a few hundred Myr,  debris would 
have spread to distances of $\sim 50$~kpc.  Estimates of the mass lost in
these events are of the order of the mass turned into stars, or roughly
$\sim 10^6$\msun.  The  average column density through this wind relic
is then only $N_H \approx 1.6 \times 10^{16} (M/10^6\msun) (50~{\rm kpc}/R)^2$.

Table~1 illustrates the equivalent widths of several strong lines.  An
H~I column of $10^{16}$~cm$^{-2}$, metallicity 1/3 solar, and solar abundance
ratios were assumed.  Regardless of whether the halo is mainly neutral H
or highly ionized, the strongest lines  would only be a few m$\AA$
wide, where $W_{\lambda} \propto N_i \lambda^2 f$. The equivalent width 
determines the S/N ratio that must be reached in the QSO continuum.  
Assuming that the continuum level itself is well-known, the S/N of the 
integral over N resolution elements is 
$S/N = \sqrt{N} \lambda / (R~ \delta W_{\lambda})$, where $R$ is
the spectral resolution $\lambda / \Delta \lambda$.  
Resolutions of 10-30 \kms
will probably be necessary to begin to resolve these lines.  Integrating
over 1 to 10 resolution elements at R = 10,000 to 30,000, the S/N must be
several hundred to detect metal lines from gas columns of order 
$10^{16}$~cm$^{-2}$. Higher columns like those detected at moderate redshift
can be found with $S/N \sim 30$.

\begin{center}
{\bf Table 1} \\
{\bf Detecting Metals in $N_{\rm HI} = 10^{16}$~cm$^{-2}$ (1/3\zsun) }
\\
  \    \\
\begin{tabular}{lll}
\hline\hline
Ion     & $W_{\lambda}(10^{16})$        & S/N   \\
        &  \angs\                       & 5$\sigma$     \\
\hline
Mg II 2796.3    & 0.0054                & 270 \\
Si II 1263.3    & 0.0017                & 400 \\
C II 1334.5     & 0.0024                & 290   \\
C IV 1549.1     & 0.0073                & 110   \\
S iIV 1396.3    & 0.0016                & 460 \\
O VI 1033.8     & 0.0053                & 100   \\
\hline \\
\end{tabular}
\end{center}

\begin{center}
{\bf Table 2} \\
{\bf Limiting QSO B Magnitude at $\sim 15$\kms Resolution$^a$}  \\
  \    \\
\begin{tabular}{lllll}
\hline\hline
Exp Time        & S/N   &       $10^4$~s        & $10^5$~  & $10^6$~s
\\
\hline
STIS G230M &30  &       14.6    & 16.9          & 18.6  \\
COS  G260M & 30 &       14.4    & 16.9          & 19.4 \\
STIS G140M &30  &       14.8    & 17.3          & 19.8  \\
COS G160M  & 30 &       14.0    & 16.5          & 19.0 \\
STIS G230M & 300&       9.8     & 12.3          & 14.8  \\
STIS G140M & 300&       9.8     & 12.3          & 14.8 \\
SUVO?      & 30 &       20      & 22.5          & 25    \\
SUVO?      & 300&       15      & 17.5          & 20    \\
\hline
\end{tabular}
\end{center}
\noindent
$^a$ An observed spectral energy distribution with
$F_{\lambda} \propto \lambda^{-1}$ was assumed.
In practice the maximum achievable $S/N$ ratio may be limited
by the exposure and stability of the flatfields. \\

Table~2 illustrates the quasar $B$ magnitude  that can  be reached  with the
current and planned instruments on HST.  The three exposure-time columns
represent practical limits for: (1) a survey project; (2) a 
special target; and (3) a campaign like the Hubble Deep Field.
Sensitivities for STIS were taken from the instrument handbook and the
on-line exposure time calculator.  The estimates for COS assume that
S/N = 10 will be obtained in $10^4$~s at $F_{\lambda} = 
1.5 \times 10^{-15}$\flux
\angs$^{-1}$.  Over a single line, the gain with COS is not
dramatic; but COS provides about 6 times more spectral coverage per tilt
at this resolution.  At low S/N, the detector noise limits the depth
reached with STIS in the near-UV, so COS will provide a dramatic gain.

If detector noise is negligible, then 
the number of potential sightlines increases by a factor of $\sim 160$
with each successive column to the right.
At $B = 18$, the average area on the sky per quasar is about 2 square degrees.  
However, because the counts rise so steeply with magnitude, one could obtain
a sightline per galaxy, about $10' \times 10'$, if $B = 20$ were accessible.
Indeed, STIS could reach $B \approx 19.8$ 
in a dedicated HDF-like campaign.  To complete 
the type of statistical study called for, however, one would like  data
for perhaps 5 sightlines towards 200 galaxies.  Without some type of
multiplexing, this program would take $\sim 10^9$~s to complete with current
and planned instruments!  Yet with only  a factor of 140 gain is sensitivity,
surveys for $N_{\rm HI} \sim 10^{17}$ cm$^{-2}$
halos could be completed in a year.
Multiplexing could reduce the required gain in sensitivity
by a factor of 5.  With this type of improvement,
Table~2 shows that it also becomes possible to observe selected lines of
sight at S/N=300 and measure the metallicity 
in columns with $N_{\rm HI} \approx 10^{16}$~cm$^{-2}$.

\subsection{The Dispersal of Metals}

Along starburst galaxy sightlines, moderately high S/N is required
to measure weaker, unsaturated lines.  Pettini \& Lipman (1995) suggest
that measurements of S~II $\lambda\lambda 1250.6, 1253.8, 1259.5$ at 
$S/N = 10$ could determine whether the H~I halo is significantly more 
metal-poor than the star-forming regions.  Measurement of the O column, 
however, will
require either much higher sensitivity to reach O~I$\lambda 1355$ or
access to O~I multiplets shortward of 1190 \angs. It is perhaps
worth emphasizing that good velocity resolution, $\sim 10$\kms,
in needed to untangle individual kinematic components which would artificially
broaden the lines.  Similar techniques will provide important measures
of abundances in intergalactic clouds.  The sum of metals inside and outside
galaxies provide an important check on the integrated cosmic star formation 
history (Renzini 1997).

\subsection{Does Ionizing Radiation Escape from Galaxies?}

The contribution of galaxies to the metagalactic ionizing radiation should
be measured locally and at high redshift.  The escape fraction may be different
at various epochs, and the source of systematic uncertainties are different
at these epochs.  It is also conceivable that the escape fraction would be
higher in starbursting dwarf galaxies, so similar measurements must eventually
be made for these galaxies.  Modeling the gas-phase absorption of the Milky
Way near the Lyman limit is complex. In practice, galaxies at $z > 0.03$
are preferred (Hurwitz \et 1997).  At $z = 2.5$, the Lyman break enters
the optical domain.  At moderate reshifts HST could, in principle, observe the 
Lyman break; but it is not sensitive below $z = 0.3$ because of the fall-off
in system throughput.

Measuring the continuum flux density below the Lyman limit is challenging
at any wavelength/redshift.  For example, a bursting dwarf galaxy
and a star-forming, spiral galaxy would produce H ionizing photons at rates
of $Q_H \approx 5 \times 10^{51}$~s$^{-1}$ and $Q_H \approx 5 \times 10^
{53}$~s$^{-1}$, respectively.  For these production rates, Figure~7 shows the 
redshifted flux density just below the Lyman limit.  Some starbursts lie above 
this line.  Leitherer \et (1995) examined four starbursts with 
$Q_H = 9.5 \times
10^{52}$~s$^{-1}$ to $3.6 \times 10^{54}$~s$^{-1}$. The rates in some
high-redshift Lyman break galaxies are even higher, and a typical 1500\angs\
flux density is shown in Figure~7.  Note that the solid lines
severely overpredict the observed flux density at $z > 3$ where the opacity
from intergalactic H~I attenuates the flux by an order of magnitude.

Figure~7 clearly shows the advantage of making measurements at 940\angs\ to
980\angs.  Even locally though, fluxes about a factor of 10 lower than those
reached with HUT are needed to measure bursting dwarfs.  To compare the 
feasibility at moderate and high-redshift, a common resolution of 
$\sim 400$\kms is used in Figure~7, and a $10^4$~sec observation is required
to reach a S/N of 10.
The points for STIS are based on the G140L and G230L gratings.  The relative
gain with COS, open symbols,  would  be dramatic if the comparison was made
to the STIS echelle modes; but an extensive spectral range is not required
for this problem.    The hexagons 
represent a single 10-m telescope with total system throughputs of 4\%
and 40\%.  The figure is overly optimistic about the ease of  detecting
the continuum in high-redshift galaxies because no intergalactic
attenuation is included in the model.  However, the star
formation rates in some high-redshift starbursts are much larger,
possibly factors $\sim 100$, than the models shown, so detections   are
likely to be made first for the high-redshift starbursts.

\hbox{ \psfig{file=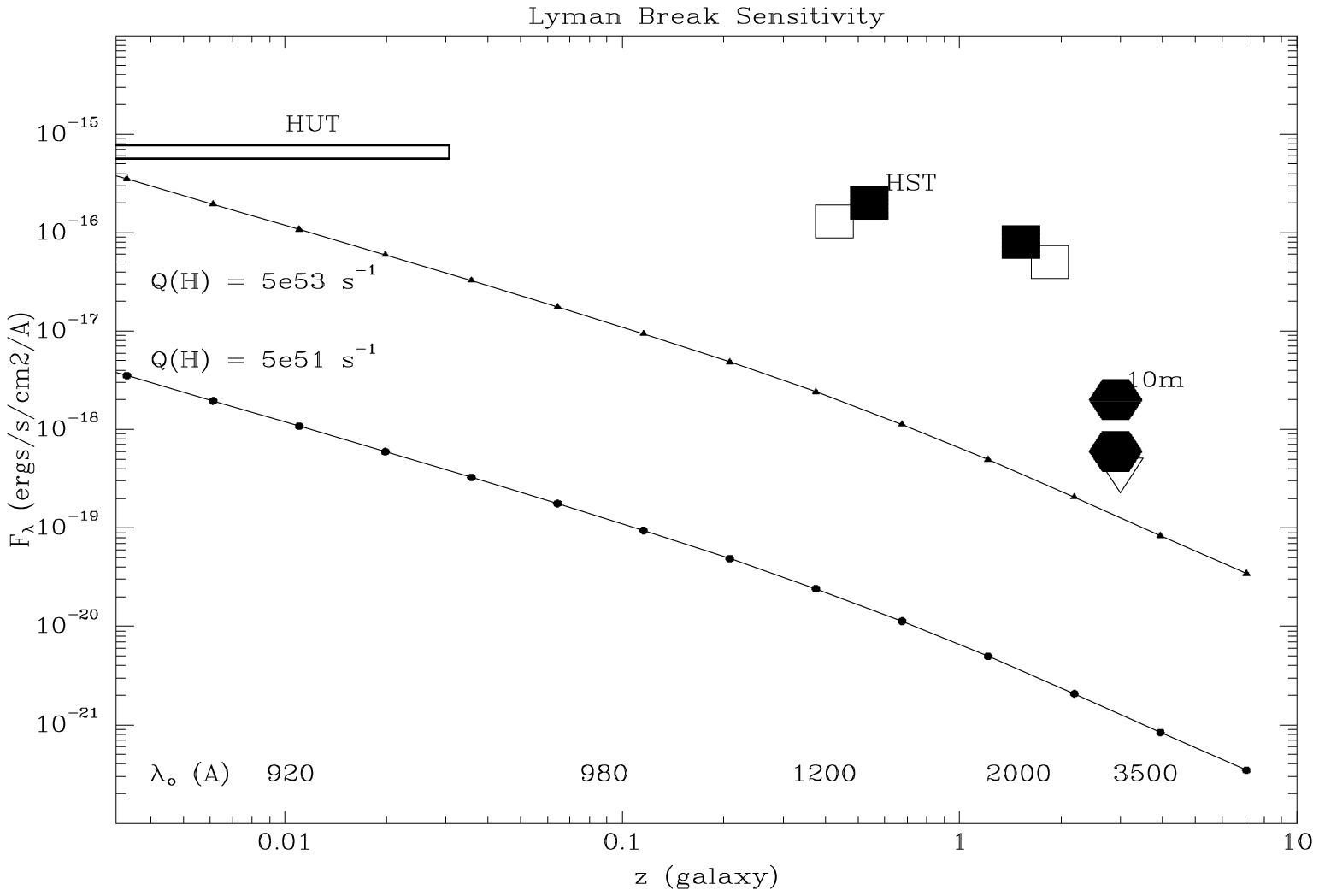,angle=0,height=3.5in} \hfill }
{Figure~7. Unattenuated flux density shortward of the Lyman break as a
function of the galaxy's redshift. Curves are shown for a 30-Doradus-like
H~II region and a star-forming, spiral galaxy.  
The inverted triangle shows the flux density at 1500\angs\ for a typical 
Lyman break galaxy  (Pettini \et 1997).  The large symbols represent
the sensitivity levels of particular instruments, see text for details.
}

\section{Conclusions}

This paper has drawn attention to several fundamental questions
about the impact of starbursts on the surrounding interstellar and
intergalactic gas.  It should be clear that answering these questions
 will require UV spectroscopy at resolutions  $\ge 10,000$ and
sensitivities approximately 140 times higher than those reachable in
surveys with HST+STIS or HST+COS.   Some type of multiplexing should be
considered to obtain spectra of several quasars  per galaxy halo
simultaneously. A strong effort should also be made to determine whether
detector senstivity can be increased by a factor $\sim 10$ in the UV.
With such advances, the aperture of the next space UV
observatory need only exceed that of HST by a factor of 3 to
provide the opportunity for extraordinary advances.

\acknowledgments
This work was supported by NASA through Hubble Fellowship grant 
HF-01083.01-96A.  It is a pleasure to thank David Bowen, 
Tim Heckman, Claus Leitherer, Max Pettini, and Daniel Weedman 
for contributing figures from their work.

\end{document}